\documentclass[12pt,a4paper,final]{iopart}

\usepackage{iopams}
\usepackage{graphicx}

\usepackage{amssymb}
\usepackage{amscd}	
\usepackage{enumerate}
\usepackage{subfigure}
\usepackage{xcolor}
\usepackage{color}
\usepackage[margin=1cm]{caption}

\newcommand{\eqref}[1]{(\ref{#1})}

\eqnobysec

\begin{document}





\title{Randomness Requirement on CHSH Bell Test in the Multiple Run Scenario}

\author{Xiao Yuan$^{*}$, Zhu Cao, Xiongfeng Ma$^{\dag}$}
\address{Center for Quantum Information, Institute for Interdisciplinary Information Sciences, Tsinghua University, Beijing 100084, China}
\ead{$^*$yuanxiao12@mails.tsinghua.edu.cn, $^\dag$xma@tsinghua.edu.cn}

\begin{abstract}
The Clauser-Horne-Shimony-Holt inequality test is widely used as a mean of invalidating the local deterministic theories and a tool of device independent quantum cryptographic tasks. There exists a randomness (freewill) loophole in the test, which is widely believed impossible to be closed perfectly. That is, certain random inputs are required for the test.  Following a randomness quantification method used in literature, we  investigate the randomness required in the test under various assumptions. By comparing the results, one can conclude that the key to make the test result reliable is to rule out correlations between multiple runs.
\end{abstract}

\vspace{2pc}
\maketitle

\section{Introduction}
Historically, Bell tests \cite{bell} are proposed for distinguishing quantum theory from local hidden variable models (LHVMs)  \cite{Einstein35}. In a general picture, a Bell  test involves multiple parties who randomly choose inputs and generate outputs with pre-shared physical resources. Based on the probability distributions of inputs and outputs, an inequality, called Bell's inequality, is defined. Any Bell test is meaningful only if all LHVMs satisfy the Bell's inequality; while in quantum mechanics, such inequality can be violated via certain quantum settings. Experimental observation of the violation of Bell's inequality would show that LHVMs are not sufficient to describe the world, and other theories, such as the quantum mechanics, are demanded.

\begin{figure*}[thb]
\centering
\resizebox{14cm}{!}{\includegraphics[scale=1]{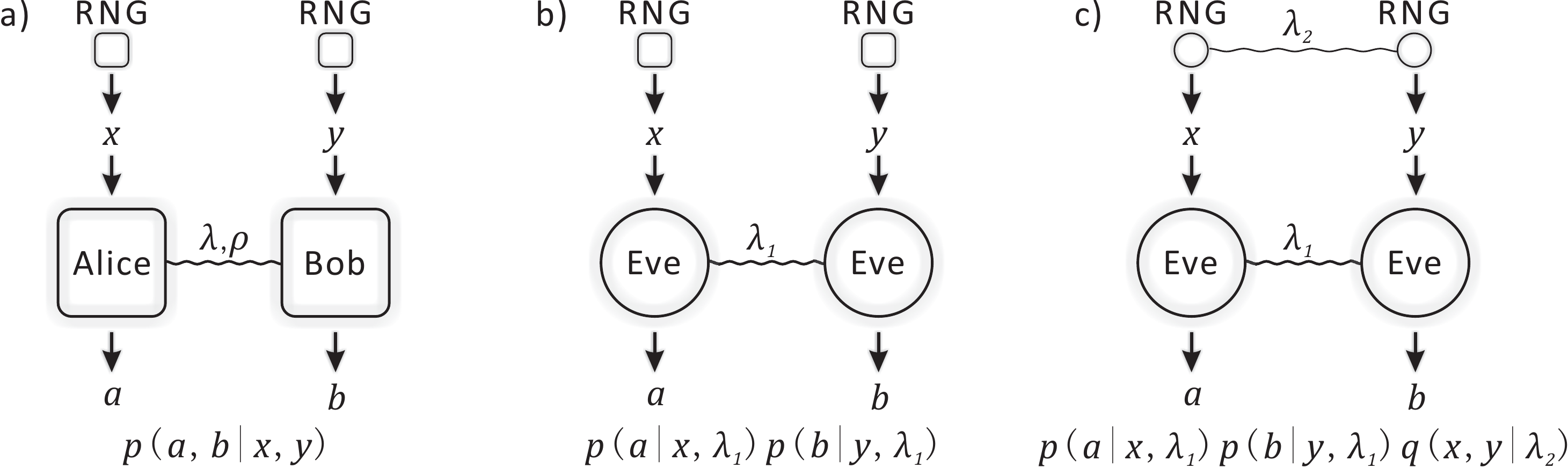}}
\caption{Bell tests in the bipartite scenario. (a) The inputs of Alice and Bob, $x$ and $y$,  are decided by perfect random number generators (RNGs), which produce uniformly distributed random numbers; (b) The measurement devices are controlled by an adversary Eve through local hidden variables $\lambda_1$; (c) The input random numbers are additionally controlled by some local hidden variable $\lambda_2$, which is accessible to Eve.}\label{Fig:BellTest}
\end{figure*}

In this work, we focus on the bipartite scenario and investigate one of the most well-known Bell tests, the Clauser-Horne-Shimony-Holt (CHSH) inequality \cite{CHSH}. As shown in Fig.~\ref{Fig:BellTest}(a), two space-like separated parties, Alice and Bob, randomly choose input settings $x$ and $y$ from an input set $I=\{0,1\}$ and generate outputs bits $a$ and $b$ based on their inputs and pre-shared quantum ($\rho$) and classical ($\lambda$) resources, respectively. The probability distribution $p(a,b|x,y)$,  obtaining outputs $a$ and $b$ conditioned on inputs $x$ and $y$, are determined by specific strategies of Alice and Bob. By assuming that the input settings $x$ and $y$ are chosen fully randomly and equally likely, we define the CHSH inequalities by convex combining the probability distribution $p(a,b|x,y)$ according to
\begin{equation}\label{eq:Bell}
  S = \sum_{a,b,x,y} (-1)^{a\oplus b + x\cdot y}p(a,b|x,y) \leq S_C = 2,
\end{equation}
where the plus operation $\oplus$ is modulo 2, $\cdot$ is numerical multiplication, and $S_C$ is the (classical) bound of Bell value $S$ for all LHVMs. Similarly, there is an achievable bound $S_Q = 2\sqrt{2}$ for the quantum theory \cite{cirel1980quantum}. In this case, a violation of the classical bound $S_C$ indicates the need for alternative theories other than LHVMs, such as quantum theory. For general no signalling (NS) theories \cite{prbox}, we denote the corresponded upper bound as $S_{NS} = 4$. It is straightforward to see that $S_{NS} \geq S_Q\geq S_C$.

In practice, the technique of violating a Bell's inequality can be applied to other quantum information tasks, such as, device independent quantum key distribution \cite{mayers1998quantum, acin06} and randomness expansion \cite{colbeck2012free, Dhara14}. Security proofs of these tasks are generally independent of the realization devices or correctness of quantum theory, but relies on violating a Bell's inequality. For instance, we can consider the devices of Alice and Bob as black boxes. In this case, we can assume, in the worse scenario, that an adversary Eve, instead of Alice and Bob, performs measurements as shown in Fig.~\ref{Fig:BellTest}(b). Because the two parties are space-like separated, the probability distribution generated in this way is always within the scope of LHVMs, that is, $p(a,b|x,y) = p(a|x,\lambda_1)p(b|y,\lambda_1)$. Therefore, Eve cannot fake a violation of any Bell tests, which intuitively explains the security of the device independent tasks.

Since the first experiment in the early 1980s \cite{Aspect1982PhysRevLett.49.91}, lots of lab demonstrations of the CHSH inequality has been presented. These experiment results show explicit violations of the LHVMs bound $S_C$, and meanwhile, suffer from a few technical and inherent loopholes, which might invalidate the conclusions. Two well-known technical obstacles are due to the locality loophole and the detection efficiency loophole, which can be closed with more delicately designed experiments and developed instruments \cite{Weihs98,Christensen13,giustina2013bell}. In contrast to the technical loopholes, there also exists an inherent loophole that cannot be closed perfectly in Bell tests --- the input settings may not be chosen randomly. In the worst case, the inputs can be all predetermined, which makes it possible to violate the Bell inequalities even with LHVMs. In this case, a witness of violating a Bell's inequality does not imply the demands for non-LHVM theories and the Bell test becomes meaningless. On the other hand, without the quantum theory or violation of Bell's inequalities, one cannot get provable randomness. Therefore, the assumption of true input randomness are indispensable in Bell tests because we cannot prove or disprove its existence.

Practically, suppose the input settings are partially controlled by an adversary Eve, who wants to convince Alice and Bob a violation of Bell's inequality with classical settings, as shown in Fig.~\ref{Fig:BellTest}(c). In this case, we model the imperfect randomness by assuming that the input settings $x$ and $y$ are chosen according to some probability distribution $q(x,y|\lambda_2)$, where $\lambda_2$ are local variables available to Eve. Now, the probability distribution $p(a,b|x,y)$ are defined by
\begin{equation}\label{eq:randomness}
p(a,b|x,y) = p_A(a|x,\lambda_1)p_B(b|y,\lambda_1)q(x,y|\lambda_2).
\end{equation}
We can therefore rewrite the CHSH inequality, given in Eq.~\eqref{eq:Bell}, as
\begin{equation}\label{Eq:BellFree}
  S = 4\sum_\lambda\sum_{a,b,x,y} (-1)^{a\oplus b + x\cdot y}p_A(a|x,\lambda)p_B(b|y,\lambda)q(x,y|\lambda)q(\lambda),
\end{equation}
where the factor $4$ is because of the average probability of choosing the input settings $x$ and $y$, $q(x,y)= \sum_\lambda q(x,y|\lambda)q(\lambda)$, is required to be $1/4$, and the hidden variables $\lambda_1$ and $\lambda_2$ are combined as $\lambda$. Notice that, in the extreme (deterministic) case where $q(x,y|\lambda) = 0$ or $1$ for all $x$, $y$, the local hidden variables $\lambda$ deterministically control the input settings. Then Eve is able to violate Bell tests to an arbitrary value with LHVMs. On the other hand, if Eve has no control of the input settings where $q(x,y|\lambda) = 1/4$ for all $x$, $y$, she cannot fake a violation at all.

Therefore, a meaningful question to ask is how one can assure that a violation of the CHSH inequality is not caused by Eve's attack on imperfect input randomness. That is, we want to know what the requirement of the input randomness is to guarantee that an observed violation truly stems from quantum effects. In the following, we first introduce the quantification of input randomness and review previous works on this question in Section \ref{Sec:Randomness}. Then we study a simplified case to gain the intuition behind Eve's optimal strategy in Section \ref{Sec:single}. Finally, we investigate the randomness requirement of the CHSH test and conclude our result in Section \ref{Sec:Result}. 


\section{Randomness Requirement}\label{Sec:Randomness}
Let us start with quantifying the input randomness. Here, we make use of the randomness parameter $P$ adopted in Ref.~\cite{Koh12} to fulfill such an attempt, other tools such as the Santha-Vazirani source \cite{santha1986generating} may work similarly. The parameter $P$ is defined to be the maximum probability of choosing the inputs conditioned on the hidden variable $\lambda$,
\begin{equation}\label{eq:randomness}
P = \max_{x,y,\lambda}q(x,y|\lambda).
\end{equation}
With this definition, the larger $P$ is, the less input randomness, the more control Eve has, and the easier for her to fake a quantum violation with LHVMs. In the CHSH test, $P$ takes values in the regime of $[1/4,1]$. When $P=1$, it represents the case that Eve has the most control of Alice and Bob's inputs, that is, the local hidden variable $\lambda$ can determine at least one set of values of $x$ and $y$. When $P = 1/4$, it corresponds to the case of complete randomness, where the adversary have no prior information on the inputs. Note that the definition of $P$ essentially follows the min-entropy, which is widely used to quantify randomness of a random variable $X$ in information theory, $H_{min}=-\log\left[\max_x prob(X=x)\right]$.

Intuitively, given complete randomness where $P = 1/4$, the value $S$ with LHVMs are bounded by $S_C$ as shown in Eq.~\eqref{eq:Bell}; while given the most dependent (on $\lambda$) randomness where $P=1$, the value $S$ with LHVMs could reach the mathematical maximum, $S_{NS}$ in the CHSH test. Then it is interesting to check the maximal $S$ value for $P\in(1/4,1)$ with LHVMs. In this work, we are interested in when the adversary can fake a quantum violation given certain randomness $P$. We thus exam the lower bound $P_Q$ of $P$ such that the Bell test result can reach the quantum bound $S_{Q}$ with an optimal LHVM. This lower bound $P_Q$ puts a minimal randomness requirement in a Bell test experiment. Only if the freedom of choosing inputs satisfies $P<P_Q$, can one claim that the Bell test is free of the randomness loophole.

Recently, lots of efforts have been spent to investigate such requirement of randomness needed to guarantee the correctness of Bell tests  \cite{Hall10,Barrett11,Hall11,Koh12,Pope13,Thinh13}. These works analyze under different conditions. One condition is about whether the input settings  are correlated or not in different runs. We call it \emph{single run}, referring to the case that the input settings are independent for different runs, and \emph{multiple run} referring to otherwise. The other condition is about whether the random inputs of Alice and Bob are correlated. Conditioned on these different assumptions of the input randomness, the lower bound $P_Q$ that allows LHVMs to saturate the quantum bound $S_Q$ in the CHSH Bell test is summarized in Table~\ref{table:Violation}.

\begin{table}[hbt]
\centering
\caption{The lower bound for randomness parameter $P$ defined in Eq~\eqref{eq:randomness} allowing the CHSH value $S$, defined in Eq.~\eqref{Eq:BellFree}, to reach the quantum bound $S_Q$ by LHVMs in the CHSH test under different conditions.}
\begin{tabular}{ccc}
  \hline
  &Alice Bob correlated& Alice Bob  uncorrelated\\
   \hline
  Single Run&0.285 \cite{Hall10,Koh12}&0.354 \cite{Koh12}\\
  Multiple Run&0.258 \cite{Pope13}&$\leq0.264$ (Our Work)\\
  \hline
\end{tabular}
\label{table:Violation}
\end{table}

In the single run scenario, the optimal strategy for Eve reaches $S = 24P-4$ and $S = 8P$ in the case that Alice's and Bob's input settings are correlated and uncorrelated, respectively \cite{Hall10,Koh12}. To achieve the maximum quantum violation $S_Q = 2\sqrt{2}$, the critical randomness requirement is shown in Table~\ref{table:Violation}. It is worth mentioning that if one has randomness $P \geq P_{NS} = 1/3$ and $P\geq P_{NS} = 1/2$ for the case of correlated and uncorrelated, respectively, Eve is able to recover arbitrary NS correlations.

In a more realistic scenario, the multiple run case, the input settings in different runs are generally correlated. Denote $N$ to be the number of test runs, $x_i$ ($y_i$) and $a_i$ ($b_i$) to be the input and output of Alice (Bob) for the $i$th run, where $i = 1, 2, \dots, N$, respectively. In the multiple run scenario, the input settings of Alice and Bob can be further correlated by
\begin{equation}\label{}
q(x_1, x_2, \dots, x_N, y_1, y_2, \dots, y_N|\lambda),
\end{equation}
Therefore, similar to the definition of Eq.~\eqref{Eq:BellFree}, we define the CHSH test in the multiple run case,
\begin{equation}\label{}
  S = \frac{4}{N}\sum_\lambda\sum_{\mathbf{a},\mathbf{b},\mathbf{x},\mathbf{y}} (-1)^{\mathbf{a}\oplus \mathbf{b} + \mathbf{x}\cdot \mathbf{y}}p_A(\mathbf{a}|\mathbf{x},\lambda)p_B(\mathbf{b}|\mathbf{y},\lambda)q(\mathbf{x},\mathbf{y}|\lambda)q(\lambda),
\end{equation}
where $\mathbf{a} = (a_1, a_2, \dots, a_N)$, $\mathbf{b} = (b_1, b_2, \dots, b_N)$, $\mathbf{x} = (x_1, x_2, \dots, x_N)$, $\mathbf{y} = (y_1, y_2, \dots, y_N)$, and $\cdot$ is inner product of two vectors, $\mathbf{x}$ and $\mathbf{y}$. Now, we can define the input randomness parameter, as an extension of Eq.~\eqref{eq:randomness},
\begin{equation}\label{eq:randomnessM}
P = \left(\max_{\mathbf{x},\mathbf{y},\lambda}q(\mathbf{x},\mathbf{y}|\lambda)\right)^{1/N}.
\end{equation}

It is obvious that the adversary is easier to fake a violation of a Bell test with LHVMs with increasing number of runs $N$. This is because the adversary can take advantage of additional dependence of the inputs in different runs. It has been shown that with randomness $P \geq P_{Q} = 0.258$, Eve is able to fake the maximum quantum violation $S_Q$ \cite{Pope13}. This result \cite{Pope13} puts a very strict requirement on the RNGs  to guarantee a faithful CHSH test.

A meaningful remaining question is thus to consider the case of multiple run but uncorrelated scenario. As all Bell experiments must run many times to sample the probability distribution, it is reasonable and also practical to consider a joint attack by Eve. On the other hand, the uncorrelated assumption is also reasonable in many realistic cases, where the experiment instruments of Alice and Bob are manufactured independently. In fact, if the inputs are determined by cosmic photons that are causally disconnected from each other, there should be no correlations between the input randomness of Alice and Bob \cite{Gallicchio14}.

Considering uncorrelated inputs of Alice and Bob,
\begin{equation}\label{Eq:Uncorrelated}
  q(x,y|\lambda) = q_A(x|\lambda)q_B(y|\lambda),
\end{equation}
we want to investigate the optimal attack with restricted randomness input $P$ in the following CHSH inequality,
\begin{equation}\label{Eq:BellFinal}
  S = \frac{4}{N}\sum_\lambda\sum_{\mathbf{a},\mathbf{b},\mathbf{x},\mathbf{y}} (-1)^{\mathbf{a}\oplus \mathbf{b} + \mathbf{x}\cdot \mathbf{y}}p_A(\mathbf{a}|\mathbf{x},\lambda)p_B(\mathbf{b}|\mathbf{y},\lambda)q_A(\mathbf{x}|\lambda)q_B(\mathbf{y}|\lambda)q(\lambda).
\end{equation}
That is, we want to maximize Eq.~\eqref{Eq:BellFinal} with the constraint of Eq.~\eqref{eq:randomnessM}. In particular, we are interested to see when this maximal value can reach $S_Q=2\sqrt2$.

\section{Single run case}\label{Sec:single}
We first review the optimal strategy in the single run scenario \cite{Koh12} to get an intuition behind the optimal attack of the adversary. Hereafter, we mainly focus on the scenario that Alice and Bob's inputs are uncorrelated as defined in Eq.~\eqref{Eq:Uncorrelated}. Thus, what we want is  to maximize  the CHSH value $S$,
\begin{equation}\label{Eq:}
  S = \sum_\lambda q(\lambda) S_\lambda,
\end{equation}
where
\begin{equation}\label{Eq:Slambda}
  S_\lambda = 4\sum_{a,b,x,y} (-1)^{a\oplus b + x\cdot y}p_A(a|x,\lambda)p_B(b|y,\lambda)q_A(x|\lambda)q_B(y|\lambda),
\end{equation}
with restricted randomness $P$, given in Eq.~\eqref{eq:randomness}.

Since any probabilistic LHVM, that is, $p_A(a|x,\lambda)p_B(b|y,\lambda)$, could be realized by a convex combination of deterministic ones \cite{Fine82}, it is therefore sufficient to only consider deterministic LHVMs. Due to the symmetric definition of the CHSH inequality, we only need to consider a specific strategy of $p_A(0|x, \lambda) = p_B(0|y, \lambda) = 1$, and $p_A(1|x, \lambda) = p_B(1|y, \lambda) = 0$ for some given $\lambda$, and all the other ones works similarly. By substituting the special strategy into Eq.~\eqref{Eq:Slambda}, we get
\begin{equation}\label{}
  S_\lambda = 4\left[q_A(0)q_B(0) + q_A(0)q_B(1) + q_A(1)q_B(0)- q_A(1)q_B(1)\right].
\end{equation}
Suppose $P_A = \max_{x,\lambda}\{q_A(x|\lambda)\}$, $P_B = \max_{y,\lambda}\{q_B(x|\lambda)\}$, and hence $P = P_AP_B$, $S_\lambda$ can be maximized to
\begin{equation}\label{}
  S_\lambda \leq  4\left[1 - 2(1-P_A)(1-P_B)\right] = 8(P_A + P_B - P) - 4.
\end{equation}
Given $P$, $S_\lambda$ is supper bounded by
\begin{equation}\label{}
  S_\lambda \leq 8P,
\end{equation}
where the equality holds when $P_B = 1/2$ and $P_A = 2P$. Thus, the optimal strategy with LHVMs is $S= 8P$.
Note that, when the input settings are fully random, $P=1/4$, the optimal strategy of LHVMs  is $S = 2$, which recovers the original LHVMs bound $S_C$. It is easy to see that, to saturate the quantum bound $S_Q = 2\sqrt{2}$, the randomness should be at least $P_Q = S_Q/8 = \sqrt{2}/4\approx0.354$, as shown in Table \ref{table:Violation}.

In the single run case, we only need to consider deterministic strategies of $p(a,b|x,y)$ due to the symmetric definition of the CHSH inequality. We also take advantage of this property in the derivation of the multiple run case. In addition, we can see that the optimal strategy of LHVMs is to choose $x$ or $y$ fully randomly and the other one as biased as possible. This biased optimal strategy is counter-intuitive since the adversary do not need to control the inputs of both parties, but only those of one party.
We show that this counter-intuitive feature does not hold in the optimal strategy in the multiple run case.

\section{Multiple run case}\label{Sec:Result}
Now we consider the multiple run scenario with uncorrelated input randomness. That is, optimizing Eve's LHVM strategy Eq.~\eqref{Eq:BellFinal} with constraints defined in Eq.~\eqref{eq:randomnessM}. Similar to the single run case, from the symmetric argument, we also only need to consider one specific value of $\lambda$ in the strategy: $p_A(0|x, \lambda) = p_B(0|y, \lambda) = 1$, and $p_A(1|x, \lambda) = p_B(1|y, \lambda) = 0$.  Given the probabilities of Alice's and Bob's inputs, $q_A(\mathbf{x}|\lambda)$, $q_B(\mathbf{y}|\lambda)$, the CHSH value for this specific $\lambda$ is given by Eq.~\eqref{Eq:BellFinal},
\begin{equation}\label{Eq:CHSHmultiple}
  S_\lambda = 4\left(1 - \frac{2}{N}\sum_{\mathbf{x},\mathbf{y}\in\{0,1\}^N}\mathbf{x}\cdot\mathbf{y}q_A(\mathbf{x}|\lambda)q_B(\mathbf{y}|\lambda)\right).
\end{equation}
Our attempt is therefore to maximize Eq.~\eqref{Eq:CHSHmultiple} with constraints
\begin{equation}\label{Eq:constraintsx}
q_A(\mathbf{x}|\lambda)q_B(\mathbf{y}|\lambda) \leq P^N,
\end{equation}
for all $q_A(\mathbf{x}|\lambda)$ and $q_B(\mathbf{y}|\lambda)$.

Since in the single run scenario, the optimal strategy requires only one party with biased conditional probability, we first analyze the case with only Alice's inputs biased and Bob's inputs uniformly distributed. Then we investigate the case where the inputs of both parties are biased. We can see that the one party biased strategy is not optimal in the multiple run case, even when $N=2$.

\subsection{One party Biased}
In the case when Eve only (partially) controls one of the inputs, say Alice's, the probability of Alice's input string $q_A(\mathbf{x}|\lambda)$ is biased and Bob's input string is uniformly distributed, that is,
\begin{equation}\label{eq:1biasedqB}
  q_B(\mathbf{y}|\lambda) = \frac{1}{2^N}.
\end{equation}
The randomness is characterized by Eq.~\eqref{eq:randomnessM}, after substituting Eq.~\eqref{eq:1biasedqB}, 
\begin{equation}\label{Eq:Pm0}
  P = \frac{P_A}{2},
\end{equation}
where $P_A$ is defined by $P_A = \max_{\lambda,\mathbf{x}}q_A(\mathbf{x}|\lambda)^{1/N}$. Then, the CHSH value, Eq.~\eqref{Eq:CHSHmultiple}, becomes
\begin{equation}\label{}
  S_\lambda = 4\left(1 - \frac{1}{N2^{N-1}}\sum_{\mathbf{x},\mathbf{y}\in\{0,1\}^N}\mathbf{x}\cdot\mathbf{y}q_A(\mathbf{x}|\lambda)\right).
\end{equation}

Denote the number of bit $1$ in an $N$ string $\mathbf{a}$ as $L_1(\mathbf{a})$. Given the number of bit $1$ in $\mathbf{x}$, $k_A = L_1(\mathbf{x})$,  we can sum over $\mathbf{y}$,
\begin{equation}\label{}
  \sum_{\mathbf{y}\in\{0,1\}^N}\mathbf{x}\cdot\mathbf{y} = \sum_{j = 1}^{k_A} 2^{N-k_A}j{k_A\choose j} = 2^{N-1}k_A,
\end{equation}
and group the summation of $\mathbf{x}$ according to $k_A$,
\begin{equation}\label{}
  S_\lambda = 4\left(1 - \frac{1}{N}\sum_{k_A= 0}^{N}\sum_{L_1(\mathbf{x}) = k_A}q_{A}(\mathbf{x}|\lambda)k_A\right),
\end{equation}
One only need to consider the LHVMs whose probabilities of $q_A(\mathbf{x}|\lambda)$ with the same $k_A$ are the same. Otherwise, we can always take an average of $q_A(\mathbf{x}|\lambda)$ with the same $k_A$ without increasing the randomness parameter $P$. Thus we can rewrite $S_\lambda$ as
\begin{equation}\label{Eq:optBiased}
  S_\lambda = 4\left(1 - \frac{1}{N}\sum_{k_A= 0}^{N}q_{k_A}(\mathbf{x}|\lambda){N\choose k_A}k_A\right),
\end{equation}
with normalization requirement
\begin{equation}\label{Eq:conBiased}
  \sum_{k_A= 0}^{N}q_{k_A}(\mathbf{x}|\lambda){N\choose k_A} = 1,
\end{equation}
and constraints defined in Eq.~\eqref{Eq:constraintsx}.


The optimization of Eq.~\eqref{Eq:optBiased} can be solved sufficiently via linear programming. Intuitively, to maximize $S_\lambda$ with given $P$ defined in Eq.~\eqref{Eq:Pm0},  we can simply assign $q_{k_A}(\mathbf{x}|\lambda)$ that has large $k_A$ be 0 and that has smaBll $k_A$  be $(2P)^N$. Suppose there exists an integer $l$ such that $P$ can be written as
\begin{equation}\label{Eq:PP}
  P = \frac{1}{2}\left[\sum_{k_A= 0}^{l}{N\choose k_A}\right]^{-1/N},
\end{equation}
then, Eq.~\eqref{Eq:optBiased} can be rewritten as
\begin{equation}\label{PP0}
  S = 4\left[1 - \frac{1}{N}\sum_{k_A= 0}^{N}\frac{1}{2}\left(\sum_{k_A= 0}^{l}{N\choose k_A}\right)^{-1/N}{N\choose k_A}k_A\right].
\end{equation}
For a general case where an integer $l$ cannot be found satisfying Eq.~\eqref{Eq:PP}, we can first find an integer $l$ such that,
\begin{equation}\label{Eq:PP1}
  \frac{1}{2}\left[\sum_{k_A= 0}^{l + 1}{N\choose k_A}\right]^{-1/N}< P \leq \frac{1}{2}\left[\sum_{k_A= 0}^{l}{N\choose k_A}\right]^{-1/N}.
\end{equation}
Then we can assign $q_{k_A}(\mathbf{x}|\lambda)$ to be
\begin{equation}\label{Eq:PP2}
  q_{k_A}(\mathbf{x}|\lambda) = \left\{
  \begin{array}{cc}
    (2P)^N & k_A\leq l \\
    \frac{\left[1 - \sum_{k_A= 0}^{l}(2P)^N{N\choose k_A}\right]^{-1/N}}{{N\choose l+1}} & k_A = l + 1 \\
    0 & k_A > l+1
  \end{array}
  \right.
\end{equation}

\begin{figure}[thb]
  \centering
  \resizebox{16cm}{!}{\includegraphics[scale=1]{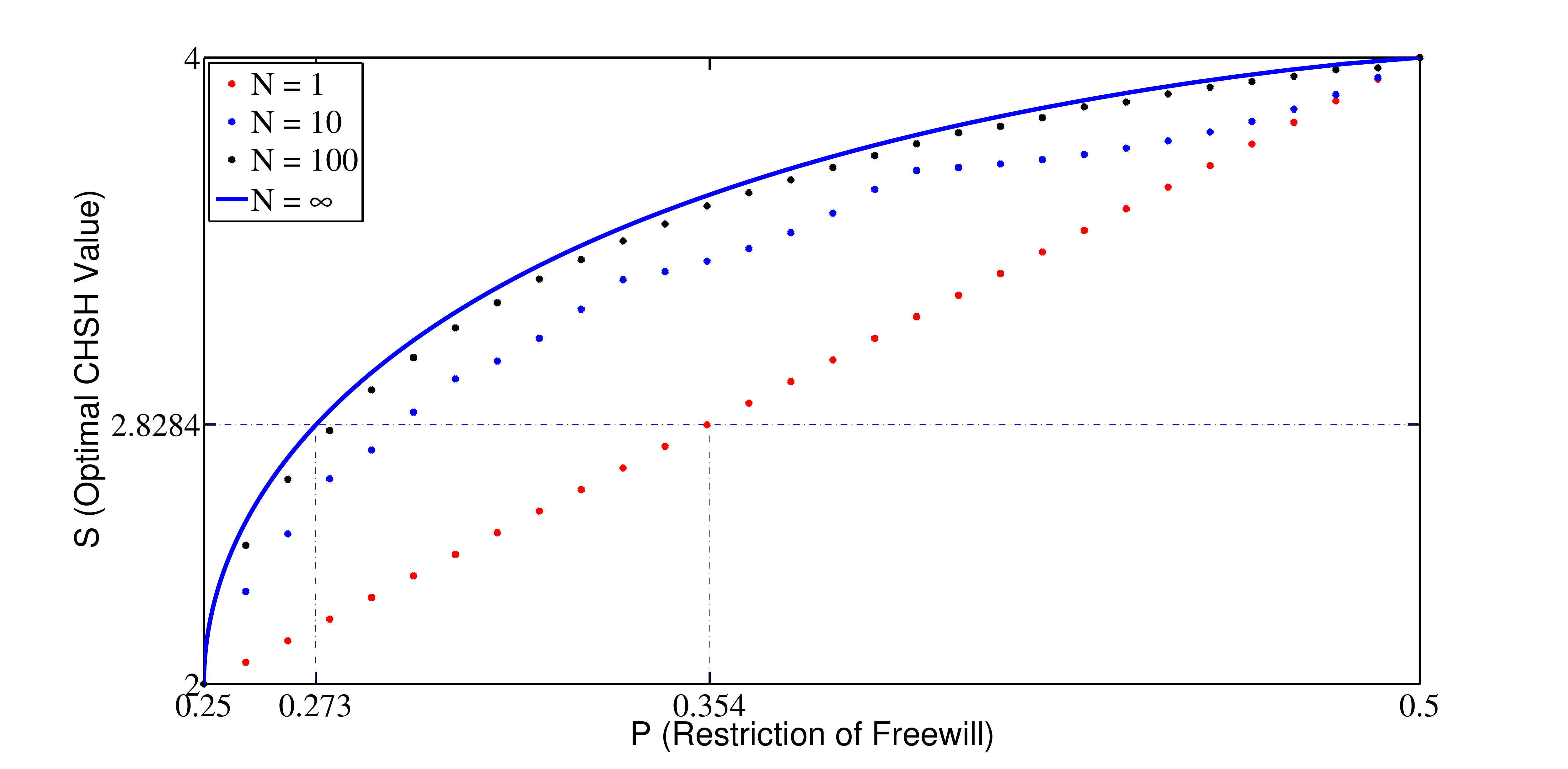}}
  \caption{Optimal value of the CHSH test for given randomness $P$ with various rounds $N$ based on only one party biased randomness. The solid line is the optimal strategy for $N\rightarrow\infty$, which upper bounds all finite $N$ rounds.  Note that the curve is not smooth for finite runs $N$ because the optimal strategy $q_{k_A}$ defined in Eq.~\eqref{Eq:PP2} jumps in $l$.  With $N$ grows larger, the curve tends to be smoother.  }\label{Fig:OneBiased}
\end{figure}

For finite $N$, one can numerically solve the problem according to Eq.~\eqref{Eq:PP2}. As shown in Fig.~\ref{Fig:OneBiased}, the optimal strategy for $N = 1, 10, 100$ are calculated. With increasing $N$, the optimal value $S$ increases and hence a valid Bell test requires a smaller $P$ (more randomness).

In the case of $N\rightarrow\infty$, we can derive an analytic bound for all finite $N$ strategies. By following the technique used in Ref.~\cite{Pope13}, we first  can estimate $P$ defined in Eq.~\eqref{Eq:PP1} with the limit of $N\rightarrow\infty$ by,
\begin{equation}\label{Eq:PBound}
  \lim_{N\rightarrow\infty}P = \frac{1}{2}\bar{l}^{\bar{l}}(1-\bar{l})^{1-\bar{l}},
\end{equation}
where $\bar{l} = l/N$, and similarly $S$ by,
\begin{equation}\label{Eq:Sinfty}
    \lim_{N\rightarrow\infty}S =  4-4\bar{l},
\end{equation}
where $\epsilon$ is an arbitrary positive number. Then we can substitute Eq.~\eqref{Eq:Sinfty} into Eq.~\eqref{Eq:PBound}, and get a relation between optimized CHSH value $S$ and the corresponding randomness parameter $P$,
\begin{equation}\label{Eq:bound}
  P=\frac{1}{2}\left(\frac{4-S}{4}\right)^{(4-S)/4}\left(\frac{S}{4}\right)^{(S/4)}.
\end{equation}

By substituting the quantum bound $S_Q = 2\sqrt{2}$ into Eq.~\eqref{Eq:bound}, we can get the critical randomness requirement to be $P_Q = 0.273$. Note that, although Eve only control Alice's input settings, she can still fake a quantum violation with sufficiently low randomness, which is lower than the single run case even when Alice's and Bob's inputs are correlated. Thus we show that the randomness is more demanded for the conditions of multiple/single run compared to the correlation between Alice and Bob.


\subsection{Both parties biased}
Now we consider a general attack, where Eve controls both inputs of Alice and Bob. In this case, we need to optimize Eq.~\eqref{Eq:CHSHmultiple} with constraints defined in Eq.~\eqref{Eq:constraintsx}.

Similarly, we also group the summation of $\mathbf{x}$ and $\mathbf{y}$ according to the corresponded number of bit $1$, $k_A = L_1(\mathbf{x})$ and $k_B = L_1(\mathbf{y})$,
\begin{equation}\label{Eq:CHSH2}
  S_\lambda = 4\left(1 - \frac{2}{N}\sum_{k_A,k_B = 0}^{N}\sum_{L_1(x) = k_A}\sum_{L_1(y) = k_B}q_{A}(\mathbf{x}|\lambda)q_{B}(\mathbf{y}|\lambda)\mathbf{x}\cdot\mathbf{y}\right).
\end{equation}
Now, if we assume that $q_{A}(\mathbf{x}|\lambda)$ ($q_{B}(\mathbf{y}|\lambda)$) has the same value for equal $k_A$ ($k_B$), we can sum over $\mathbf{x}$ and $\mathbf{y}$  for given $k_A$ and $k_B$,
\begin{eqnarray}
  \sum_{k_A,k_B}\mathbf{x}\cdot\mathbf{y} &=& {N \choose k_A}\sum_{j = \max\{1, k_A + k_B - N\}}^{\min\{k_A,k_B\}}j{k_A \choose j}{N - k_A \choose k_B - j}\nonumber \\
  &=& {N \choose k_A}k_A{N-1 \choose k_B-1}\nonumber \\
    &=& \frac{k_Ak_B}{N}{N \choose k_A}{N \choose k_B}.
\end{eqnarray}
We can then get the $S$ value to be
\begin{equation}\label{Eq:slambda}
  S_\lambda = 4\left(1 - \frac{2}{N^2}\sum_{k_A,k_B = 0}^{N}q_{k_A}(\mathbf{x}|\lambda){N \choose k_A}q_{k_B}(\mathbf{y}|\lambda){N \choose k_B}{k_Ak_B}\right),
\end{equation}
with the constraints of $q_A(\mathbf{x}|\lambda)$ and $q_B(\mathbf{y}|\lambda)$,
\begin{eqnarray}\label{Eq:constrainty}
\sum_{k_A = 1}^{N} q_{k_A}(\mathbf{x}|\lambda){N\choose k_A} &=& 1,\nonumber \\
\sum_{k_B = 1}^{N} q_{k_B}(\mathbf{y}|\lambda){N\choose k_B} &=& 1.
\end{eqnarray}

It is worth mentioning that the assumption that $q_{A}(\mathbf{x}|\lambda)$ ($q_{B}(\mathbf{y}|\lambda)$) takes the same value for equal $k_A$ ($k_B$) is not obviously equivalent to the original optimization problem defined in Eq.~\eqref{Eq:CHSH2}. We thus take this step as an additional assumption, and conjecture it to be true for certain cases of the input randomness.

The problem defined in Eq.~\eqref{Eq:slambda} with constraints of Eq.~\eqref{Eq:constrainty} cannot be solved by linear programming directly, as to the nonlinear terms $q_{k_A}(\mathbf{x}|\lambda)q_{k_B}(\mathbf{y}|\lambda)$.  However, we can still optimize it with similar methods used in the previous section.
 Define the maximum randomness on each side
\begin{eqnarray}\label{}
    P_A &=& [\max_{\lambda,\mathbf{x}}q_{k_A}(\mathbf{x}|\lambda)]^{1/N}, \nonumber\\
    P_B &=& [\max_{\lambda,\mathbf{y}} q_{k_B}(\mathbf{y}|\lambda)]^{1/N}.
\end{eqnarray}
To maximize $S_\lambda$, we can do it first for the Alice side, $q_{k_A}$, and then Bob side $q_{k_B}$. By doing so, it is not hard to see that $S_\lambda$ is maximized by assigning $q_{k_A}$ that has small number of $k_A$ to be $P_A$ and that has large number of $k_A$ to be 0, and similarly for $q_{k_B}$.
Thus we need to first find $l_A$ and $l_B$ for Alice and Bob, such that,
\begin{eqnarray}\label{Eq:PAB}
  \left[\sum_{k_A= 0}^{l_A + 1}{N\choose k_A}\right]^{-1/N}&< &P_A \leq \left[\sum_{k_A= 0}^{l_A}{N\choose k_A}\right]^{-1/N}\nonumber\\
    \left[\sum_{k_B= 0}^{l_B + 1}{N\choose k_B}\right]^{-1/N}&< &P_B \leq \left[\sum_{k_B= 0}^{l_B}{N\choose k_B}\right]^{-1/N}.
\end{eqnarray}
Then we can assign $q_{k_A}(\mathbf{x}|\lambda)$ and $q_{k_B}(\mathbf{y}|\lambda)$ to be
\begin{eqnarray}\label{Eq:qqqq}
  q_{k_A}(\mathbf{x}|\lambda) &=& \left\{
  \begin{array}{cc}
    (P_A)^N & k_A\leq l_A \\
    \frac{\left[1 - \sum_{k_A= 0}^{l_A}P_A^N{N\choose k_A}\right]^{-1/N}}{{N\choose l_A+1}} & k_A = l_A + 1 \\
    0 & k_A > l_A+1
  \end{array}
  \right.\nonumber\\
      q_{k_B}(\mathbf{y}|\lambda) &=& \left\{
  \begin{array}{cc}
    (P_B)^N & k_B\leq l_B \\
    \frac{\left[1 - \sum_{k_B= 0}^{l_B}P_B^N{N\choose k_B}\right]^{-1/N}}{{N\choose l_B+1}} & k_B = l_B + 1 \\
    0 & k_B > l_B+1
  \end{array}
  \right.
\end{eqnarray}
to optimize $S_\lambda$ defined in Eq.~\eqref{Eq:slambda}.

\begin{figure}[htb]
  \centering
  \resizebox{16cm}{!}{\includegraphics[scale=1]{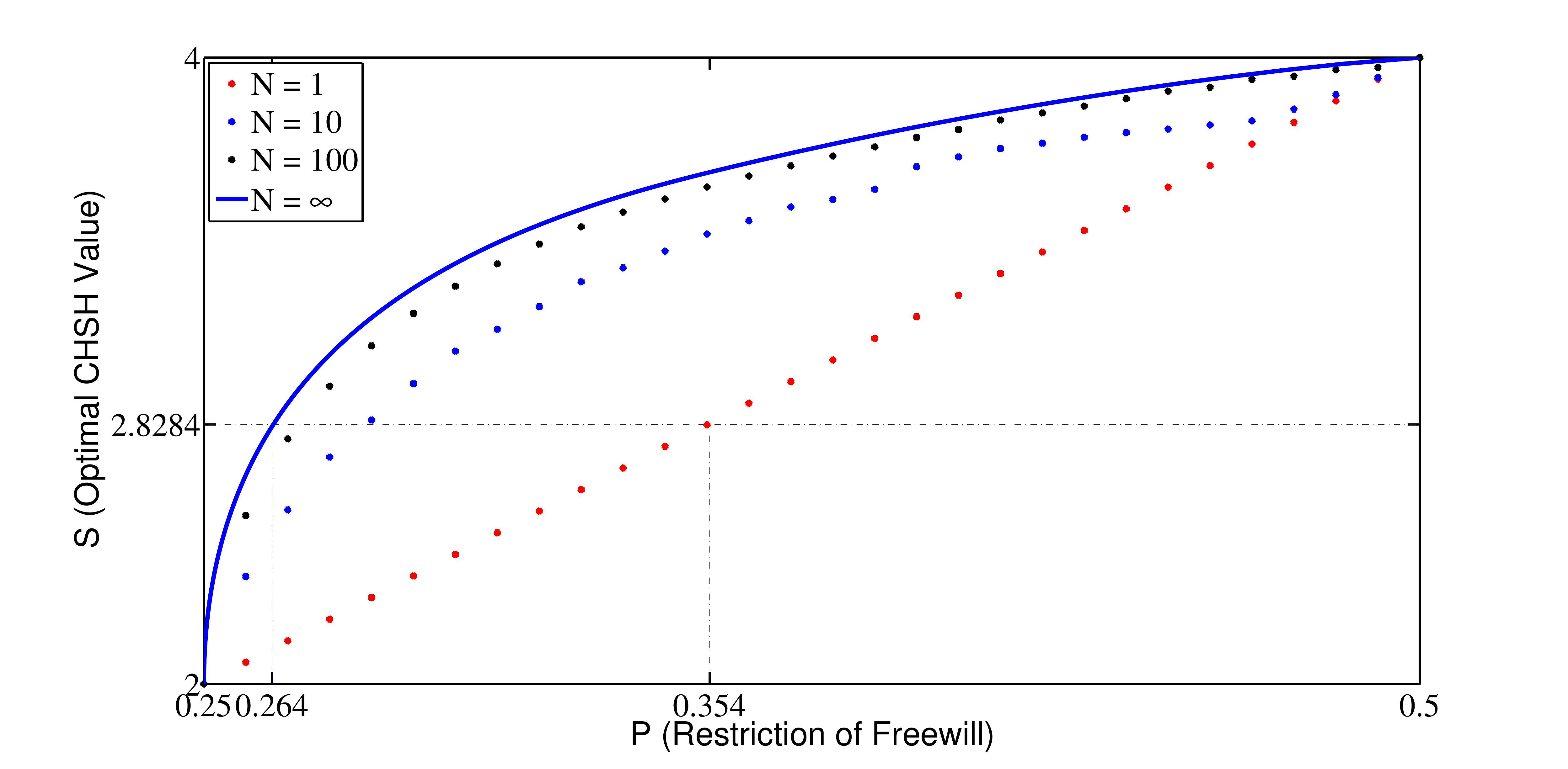}}\\
  \caption{Possible optimal value of the CHSH test for given randomness $P$ with various rounds $N$ based on uncorrelated input. The solid line corresponds the strategy for $N\rightarrow\infty$, which upper bounds all finite $N$ cases. The curves are not smooth for finite $N$ as for similar reasons like in the one party biased case, and it tends to be smooth with $N\rightarrow\infty$.}\label{Fig:Biased}
\end{figure}

For finite $N$, we can also numerically solve the optimization problem defined in Eq.~\eqref{Eq:slambda}. As shown in Fig.~\ref{Fig:Biased}. The value $S$ increases with the number of runs $N$, thus the strategy with infinite rounds puts a bound on the strategy with finite rounds.

In the case of  $N\rightarrow\infty$, we can also find analytical relation between optimized $S$ and the corresponded $P$. Similarly, we can first estimate $P_A$ and $P_B$ defined in Eq.~\eqref{Eq:PAB} with the limit of $N\rightarrow\infty$  by
\begin{eqnarray}\label{}
    \lim_{N\rightarrow\infty} P_A &=  & \bar{l}_A^{\bar{l}_A}(1-\bar{l}_A)^{1-\bar{l}_A}, \nonumber\\
    \lim_{N\rightarrow\infty} P_B &=  & \bar{l}_B^{\bar{l}_B}(1-\bar{l}_B)^{1-\bar{l}_B},
\end{eqnarray}
where $\bar{l}_A = l_A/N$ and $\bar{l}_B = l_B/N$, and $S$ according to
\begin{equation}\label{}
    S =  4 - 8\bar{l}_A\bar{l}_B.
\end{equation}

As we still have to optimize over all possible $P_A$ and $P_B$ that satisfies $P_AP_B = P$, we cannot get a direct analytic formula like in Eq.~\eqref{Eq:bound}, while we can still numerically solve  and plot it in Fig.~\ref{Fig:Biased}.
To reach a maximum quantum violation $S_Q = 2\sqrt{2}$ with a LHVM, the randomness is required to be $P\geq P_Q \approx 0.264$.

\section{Discussion}
We take an additional assumption in the derivation of the both parties biased case, thus the obtained bound $P_Q \approx 0.264$ is still an upper bound of a general optimal attack. As we already know, the randomness requirement for the worst case, that is, multiple run with Alice and Bob correlated, is strictly bounded by $P_Q \approx 0.258$ \cite{Pope13}. Thus, we know that the tight $P_Q$ for the case of multiple run but Alice and Bob uncorrelated should lie in the regime of $[0.258, 0.264]$.

To gain intuition why we take the additional assumption, first notice that what we want is to minimize the average contribution of $\mathbf{x}\cdot \mathbf{y}$ in Eq.~\eqref{Eq:CHSH2}. 
In our case, where $P$ is near 1/4, $q_A(\mathbf{x|}\lambda)$ and $q_B(\mathbf{y|}\lambda)$ can be regarded as an approximately flat distribution.
On average, the $\mathbf{x}$ ($\mathbf{y}$) contains less number of 1s will contribute more to $S$, which means we should assign the corresponded probability $q_A(\mathbf{x|}\lambda)$ ($q_A(\mathbf{y|}\lambda)$) bigger in order to maximize $S$. As $q_A(\mathbf{x|}\lambda)$ ($q_A(\mathbf{y|}\lambda)$) is upper bounded by $P_A$ ($P_B$), an intuitive optimal strategy is then to let $q_A(\mathbf{x|}\lambda)$ ($q_A(\mathbf{y|}\lambda)$) be $P_A$ ($P_B$) for $\mathbf{x}$ ($\mathbf{y}$) contains less number of 1s, and be 0 for the ones contains more number of 1s. As $q_A(\mathbf{x|}\lambda)$ ($q_A(\mathbf{y|}\lambda)$) should also satisfy the normalization condition (Eq.~\eqref{Eq:constrainty}), we can simply follow the strategy defined in Eq.~\eqref{Eq:qqqq} to realize the intuition, which on the other hand satisfies the assumption we take.
Follow the above intuition, we conjecture the assumption to be true for certain cases of $N$.  That is, for finite $N$, we conjecture it to be true when equalities are taken in Eq.~\eqref{Eq:PAB} for both $P_A$ and $P_B$.

On the other hand, we want to emphasize that for a finite $N$, the assumption will not generally hold in the optimal strategy if the equalities in Eq.~\eqref{Eq:PAB} are not fulfilled. For example, if the probability of $l_A+1$ and $l_B+1$  in Eq.~\eqref{Eq:qqqq} is not 0 but very small, we should not take all $q_A(\mathbf{x}|\lambda)$ and $q_B(\mathbf{y}|\lambda)$ equally as $q_{k_A}$ and $q_{k_B}$, especially for the case of
$L_1(\mathbf{x}) = l_A+1$ and $L_1(\mathbf{y}) = l_B+1$ , respectively. In fact, there do exists a cleverer  assignment of $q_A(\mathbf{x}|\lambda)$ and $q_B(\mathbf{y}|\lambda)$ such that only $\mathbf{x}$ and $\mathbf{y}$ that gives small $\mathbf{x}\cdot\mathbf{y}$ get probability instead of all of $\mathbf{x}$ and $\mathbf{y}$ that $L_1(\mathbf{x}) = l_A+1$ and $L_1(\mathbf{B}) = l_B+1$.
However, with increasing runs $N$, this kind of clever attack stops working as for the equalities can be more approximately satisfied with larger $N$. Therefore, we also conjecture the assumption to be true for all possible $P$ with $N$ goes to infinity.

As we can see, our obtained $P_Q\approx 0.264$ is already very close to the worst case value that is $0.258$, we can therefore conclude that the multiple run correlation is already a strong resource for the adversary, no matter whether Alice and Bob are correlated or not.
In addition, as we know that the bound $P_Q$ for the most loose case, that is, single run and Alice Bob uncorrelated, is given to be $0.354$ \cite{Koh12}, we also suggest that the key loophole of the input randomness is the correlation between multiple runs instead of correlation of Alice and Bob.

\section{Conclusion}
In this work, we consider the randomness requirement of CHSH test in the multiple run scenario. By considering an adversary Eve who independently controls the input randomness of Alice and Bob, we investigate the minimum randomness requirement to guarantee that a violation of the CHSH inequality is not due to Eve's attack (LHVM).

Considering that Eve controls only Alice's input but leaves Bob's input uniformly distributed, we found the randomness Eve need to control to fake a quantum violation is $P_Q = 0.273$. And the randomness  required when controlling both Alice and Bob is $P_Q \le 0.264$. By comparing the results to the ones listed in Table.~\ref{table:Violation}, we conclude that the key randomness loophole is due to the correlation between multiple runs. As the randomness requirement which considers multiple run attack is not easy to realize in real experiments, we thus suggest the experiments to rule correlations of the input settings from different runs. To guarantee the securities of the device independent tasks, we also suggest that one should check whether there is correlation between random inputs from different runs.

For further research, we are interested to know whether there exists Bell inequalities that suffers less from the randomness loophole. By assuming different kinds of assumptions, the randomness requirement behaves different. Recently, by considering a nonzero lower bound for the input random probability $p(x,y|\lambda)$, $\mathrm{P\ddot{u}tz}$ et al. show a Bell inequality which suffers from very little randomness loophole \cite{putz14}. That is, any adversary cannot fake a quantum violation as long as the lower bound of $p(x,y|\lambda)$ is nonzero regardless of its upper bound $P$ defined in Eq.~\eqref{eq:randomness}. Therefore, it is interesting to investigate the multiple run randomness requirement of the CHSH inequality with additional assumption.


\section*{Acknowledgments}
The authors acknowledge insightful discussions with S.~Yang, Q.~Zhao, and Z.~Zhang. This work was supported by the National Basic Research Program of China Grants No.~2011CBA00300 and No.~2011CBA00301, and the 1000 Youth Fellowship program in China.


\section*{References}

\bibliographystyle{unsrt}

\bibliography{BibBell}

\begin{thebibliography}{10}

\bibitem{bell}
John~Stuart Bell.
\newblock {\em On the Einstein-Podolsky-Rosen Paradox. Physics 1, 195--200
  (1964)}.
\newblock Speakable and Unspeakable in Quantum Mechanics. Cambridge University
  Press, 1987.

\bibitem{Einstein35}
A.~Einstein, B.~Podolsky, and N.~Rosen.
\newblock Can quantum-mechanical description of physical reality be considered
  complete?
\newblock {\em Phys. Rev.}, 47:777--780, May 1935.

\bibitem{CHSH}
John~F. Clauser, Michael~A. Horne, Abner Shimony, and Richard~A. Holt.
\newblock Proposed experiment to test local hidden-variable theories.
\newblock {\em Phys. Rev. Lett.}, 23:880--884, Oct 1969.

\bibitem{cirel1980quantum}
Boris~S Cirel'son.
\newblock Quantum generalizations of bell's inequality.
\newblock {\em Letters in Mathematical Physics}, 4(2):93--100, 1980.

\bibitem{prbox}
Sandu Popescu and Daniel Rohrlich.
\newblock Quantum nonlocality as an axiom.
\newblock {\em Foundations of Physics}, 24:379--385, 1994.
\newblock 10.1007/BF02058098.

\bibitem{mayers1998quantum}
Dominic Mayers and Andrew Yao.
\newblock Quantum cryptography with imperfect apparatus.
\newblock In {\em Foundations of Computer Science, 1998. Proceedings. 39th
  Annual Symposium on}, pages 503--509. IEEE, 1998.

\bibitem{acin06}
Antonio Ac\'in, Nicolas Gisin, and Lluis Masanes.
\newblock From bell’s theorem to secure quantum key distribution.
\newblock {\em Phys. Rev. Lett.}, 97:120405, Sep 2006.

\bibitem{colbeck2012free}
Roger Colbeck and Renato Renner.
\newblock Free randomness can be amplified.
\newblock {\em Nature Physics}, 8(6):450--453, 2012.

\bibitem{Dhara14}
Chirag Dhara, Gonzalo de~la Torre, and Antonio Ac\'in.
\newblock Can observed randomness be certified to be fully intrinsic?
\newblock {\em Phys. Rev. Lett.}, 112:100402, Mar 2014.

\bibitem{Aspect1982PhysRevLett.49.91}
Alain Aspect, Philippe Grangier, and G\'erard Roger.
\newblock Experimental realization of einstein-podolsky-rosen-bohm
  <i>gedankenexperiment</i>: A new violation of bell's inequalities.
\newblock {\em Phys. Rev. Lett.}, 49:91--94, Jul 1982.

\bibitem{Weihs98}
Gregor Weihs, Thomas Jennewein, Christoph Simon, Harald Weinfurter, and Anton
  Zeilinger.
\newblock Violation of bell's inequality under strict einstein locality
  conditions.
\newblock {\em Phys. Rev. Lett.}, 81:5039--5043, Dec 1998.

\bibitem{Christensen13}
B.~G. Christensen, K.~T. McCusker, J.~B. Altepeter, B.~Calkins, T.~Gerrits,
  A.~E. Lita, A.~Miller, L.~K. Shalm, Y.~Zhang, S.~W. Nam, N.~Brunner, C.~C.~W.
  Lim, N.~Gisin, and P.~G. Kwiat.
\newblock Detection-loophole-free test of quantum nonlocality, and
  applications.
\newblock {\em Phys. Rev. Lett.}, 111:130406, Sep 2013.

\bibitem{giustina2013bell}
Marissa Giustina, Alexandra Mech, Sven Ramelow, Bernhard Wittmann, Johannes
  Kofler, J{\"o}rn Beyer, Adriana Lita, Brice Calkins, Thomas Gerrits, Sae~Woo
  Nam, et~al.
\newblock Bell violation using entangled photons without the fair-sampling
  assumption.
\newblock {\em Nature}, 497(7448):227--230, 2013.

\bibitem{Koh12}
Dax~Enshan Koh, Michael J.~W. Hall, Setiawan, James~E. Pope, Chiara Marletto,
  Alastair Kay, Valerio Scarani, and Artur Ekert.
\newblock Effects of reduced measurement independence on bell-based randomness
  expansion.
\newblock {\em Phys. Rev. Lett.}, 109:160404, Oct 2012.

\bibitem{santha1986generating}
Miklos Santha and Umesh~V Vazirani.
\newblock Generating quasi-random sequences from semi-random sources.
\newblock {\em Journal of Computer and System Sciences}, 33(1):75--87, 1986.

\bibitem{Hall10}
Michael J.~W. Hall.
\newblock Local deterministic model of singlet state correlations based on
  relaxing measurement independence.
\newblock {\em Phys. Rev. Lett.}, 105:250404, Dec 2010.

\bibitem{Barrett11}
Jonathan Barrett and Nicolas Gisin.
\newblock How much measurement independence is needed to demonstrate
  nonlocality?
\newblock {\em Phys. Rev. Lett.}, 106:100406, Mar 2011.

\bibitem{Hall11}
Michael J.~W. Hall.
\newblock Relaxed bell inequalities and kochen-specker theorems.
\newblock {\em Phys. Rev. A}, 84:022102, Aug 2011.

\bibitem{Pope13}
James~E. Pope and Alastair Kay.
\newblock Limited measurement dependence in multiple runs of a bell test.
\newblock {\em Phys. Rev. A}, 88:032110, Sep 2013.

\bibitem{Thinh13}
Le~Phuc Thinh, Lana Sheridan, and Valerio Scarani.
\newblock Bell tests with min-entropy sources.
\newblock {\em Phys. Rev. A}, 87:062121, Jun 2013.

\bibitem{Gallicchio14}
Jason Gallicchio, Andrew~S. Friedman, and David~I. Kaiser.
\newblock Testing bell’s inequality with cosmic photons: Closing the
  setting-independence loophole.
\newblock {\em Phys. Rev. Lett.}, 112:110405, Mar 2014.

\bibitem{Fine82}
Arthur Fine.
\newblock Hidden variables, joint probability, and the bell inequalities.
\newblock {\em Phys. Rev. Lett.}, 48:291--295, Feb 1982.

\bibitem{putz14}
G.~{P{\"u}tz}, D.~{Rosset}, T.~J. {Barnea}, Y.-C. {Liang}, and N.~{Gisin}.
\newblock {Quantum nonlocality beats arbitrary lack of free choice}.
\newblock {\em ArXiv e-prints}, July 2014.

\end{thebibliography}


\end{document}